%% file: main.tex
\documentclass[conference,10pt,a4paper]{IEEEtran}

\usepackage{silence}
\newlength{\flexwidth}
\usepackage{bm}
\usepackage{amsmath,amssymb,amsthm,fixmath}
\usepackage{mathtools}
\usepackage{accents}
\usepackage{color,colortbl}
\usepackage{xcolor}
\usepackage{siunitx}
\usepackage{cuted}
\usepackage{multirow,booktabs}
\usepackage{subcaption}
\usepackage[inline]{enumitem}
\usepackage{optidef}
\usepackage{graphicx}
\usepackage{lipsum}

\usepackage{amsmath}
\usepackage{epstopdf}
\usepackage[normalem]{ulem}
\newcommand{\revise}[2]{{\color{red}\sout{#1}}{\color{blue}#2}}
\renewcommand{\revise}[2]{#2}
\usepackage[textsize=tiny,colorinlistoftodos]{todonotes}
\makeatletter
\define@key{todonotes}{bh}[]{
	\setkeys{todonotes}{author=\textbf{Bin}, color=lime!30}}%
\define@key{todonotes}{zf}[]{
	\setkeys{todonotes}{author=\textbf{Zexin}, color=blue!30}}%
\makeatother

\newif\ifreviewmode
\reviewmodetrue 

\ifreviewmode
\else
  \renewcommand{\todo}[1]{} 
  \renewcommand{\revise}[2]{#2} 
\fi

\usepackage[shortcuts,acronym,automake]{glossaries}
\input{glossary}

\usepackage{tcolorbox}
\tcbuselibrary{many}

\newtcbtheorem[]{alg}{Algorithm}{fonttitle=\bfseries}{alg}

\usepackage[norelsize,linesnumbered,ruled]{algorithm2e}
\SetKwRepeat{Do}{do}{while}
\makeatletter
\newcommand{\removelatexerror} {\let\@latex@error\@gobble}
\makeatother

\begin{document}
	
	\title{Unauthorized Radio Sensing and Privacy Risks: A Sampling Error-Based Defense}
	
	\author{
        \IEEEauthorblockN{Zexin~Fang\IEEEauthorrefmark{1},~Bin~Han\IEEEauthorrefmark{1},~Wenwen~Chen\IEEEauthorrefmark{1},~and~Hans~D.~Schotten\IEEEauthorrefmark{1}\IEEEauthorrefmark{2}}
		\IEEEauthorblockA{
  \IEEEauthorrefmark{1}{University of Kaiserslautern-Landau (RPTU), Germany}\\
  \IEEEauthorrefmark{2}{German Research Center for Artificial Intelligence (DFKI), Germany}
		}
	}
	
	\bstctlcite{IEEEexample:BSTcontrol}
	
	\maketitle

	\begin{abstract}  
        Unauthorized sensing activities pose an increasing threat to individual privacy, yet effective countermeasures remain underdeveloped. This paper presents a novel methodology to characterize and counter such unauthorized surveillance. We model pedestrian trajectories as a random process and leverage the \gls{crlb} to evaluate sensing performance, interpreting it as sampling error within this random process. Through simulation, we verify our method's accuracy in monitoring unauthorized sensing activities in urban environments and validate the effectiveness of our proposed mitigation strategies.
	\end{abstract}
	
	\begin{IEEEkeywords} 
	Privacy, radio sensing, tracking, \gls{crlb}.
	\end{IEEEkeywords}
	
    \IEEEpeerreviewmaketitle
    \glsresetall
    \revise{}
    With the rapid advancement of sensing and localization, environmental modeling, and data analysis techniques, our world is becoming increasingly monitored and interconnected. While these technological developments offer significant benefits for automation, safety, and efficiency across various domains \cite{safety2017Huang,Auto2019Hassan}, recent research has systematically unveiled the urgent privacy risks posed by high-accuracy radio positioning and sensing, as comprehensively documented in \cite{risk2024Nguyen}. The primary distinction between radio sensing and positioning lies in the latter requiring targets to carry cellular devices. Notably, malicious actors often combine both techniques to achieve high-accuracy tracking of unwitting targets without their consent. As privacy risks in high-accuracy radio positioning gain increasing attention, several efforts have been undertaken to address these concerns. Some researchers have focused on limiting the linkability of data \cite{traj2023guan}, while others have developed methods for encrypting Channel State Information (CSI) to prevent unauthorized CSI-based localization \cite{wsn2021Harn}. 
    
    However, the critical risk of data acquisition through radio sensing remains largely unexplored. Meanwhile, regulatory frameworks addressing this issue are currently absent, they are urgently needed at both international and national levels \cite{laws2021mani}. Legitimate sensing can embed authentication watermarks in waveforms upon certification by authorized organizations, following approaches like \cite{watermark2018Ferd}, while unauthorized sensing requires detection and monitoring. To address this, we present a novel methodology based on sensing accuracy estimation. We model pedestrian trajectories as a random process, with unauthorized sensing acting as sampling of this process. When combined with environmental modeling, this sampling enables malicious actors to analyze individual behaviors. Significant sampling errors typically result in misinterpretation of individual behaviors. Therefore, the key to preserving individual privacy against unauthorized sensing lies in monitoring sampling errors and interfering with sensing systems when necessary.

    The sampling error of a sensing system depends on waveforms, channel conditions, and sensing initiator configurations, making it challenging to estimate the actual sensing performance from the target's perspective. Instead, we estimate the achievable minimum sampling error in unauthorized sensing based on \gls{crlb}. In Section \ref{sec:sysmodel}, we describe all potential sensing errors. In particular, in Subsec.~\ref{subsec:perlow}, we consolidate these errors into a single metric to characterize the tracking performance of malicious actors. In Sec. \ref{sec:Meth}, we assume that the target carries a mobile device equipped with \gls{doa} estimation, enabling the target to obtain sampling error of sensing while intermittently jamming the sensing source. Consequently, in Sec.~\ref{sec:simu}, we conduct simulations in practical scenarios to validate the feasibility of accessing the sampling error. Based on the accessed sampling error, we propose and validate corresponding mitigation strategies. Finally, in Sec.~\ref{sec:conclu}, we conclude this paper. 

   \section{Sampling error modelling}\label{sec:sysmodel}
   \subsection{\gls{crlb} of sensing resolution}\label{subsec:descipCRLB}
   In radar waveform design, the \gls{af} specifies the output of a matched filter in the absence of noise. It is generally used to evaluate the performance of radar waveforms. The \gls{af} of a discrete signal $s[n]$ with length $N$ can be described, 
   \begin{equation}\label{eq:afd}
   \mathcal{A}\left(k,f_v\right) = \sum\limits_{n=1}^N\left\vert s[n]s^*[n+k]e^{j2\pi\frac{f_vn}{NT}}\right\vert,
	\end{equation}
   where $f_v,k$ correspond to the Doppler shift and bit time delay, and $T$ represents the sample interval. A common method for estimating radar waveform sensing resolution involves determining the point where power drops by $3$ dB from its peak in the \gls{af}. However, this approach is computationally intensive for constant monitoring; therefore, we directly access the \gls{crlb} of sensing pulses. The \gls{fim}, which quantifies parameter estimation precision from observations, is given in \cite{FIMCXTXB2012}:
   \begin{equation}\label{eq:fimafd}
    \mathbf{J}_{M}(k, f_v) =  -2\gamma
    \begin{bmatrix}
        \frac{\partial^2 \mathcal{A}\left(k,f_v\right)}{\partial^2 k} & \frac{\partial^2 \mathcal{A}\left(k,f_v\right)}{\partial k \partial f_v},\\
        \frac{\partial^2 \mathcal{A}\left(k,f_v\right)}{\partial k \partial f_v} & \frac{\partial^2 \mathcal{A}\left(k,f_v\right)}{\partial^2 f_v}
    \end{bmatrix}_{\substack{k=0\\f_v=0}},
	\end{equation}
   where $\gamma$ is the \gls{snr} of sensing pulses. 
    \begin{figure}[!htbp]
    \centering
   \includegraphics[width=0.3\textwidth]{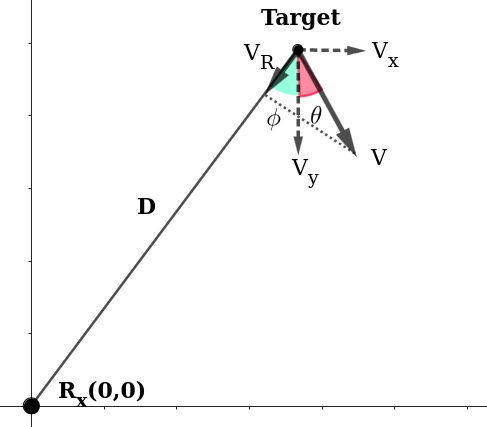}
   \caption{Monostatic sensing geometry}\label{fig:monostatic}
   \end{figure}
   In a monostatic sensing configuration, depicted in Fig.~\ref{fig:monostatic}, the delay of signal and frequency shift can be then described as follows:   
   \begin{align}
   k &= \frac{2D}{c},\label{eq:timeshiftdis}\\   
   f_v &= \frac{2f_c\cos(\phi+\theta) V}{c},\label{eq:velshiftdis}
   \end{align} 
   where $D$ is the distance between the sensing initiator and the target, $c$ denotes the speed of light, $f_s$ represents the sampling frequency, and $f_c$ is the carrier frequency of the sensing signal. Additionally, $V$ is the velocity of the target, $\phi$ is the look angle from the receiver, and $\theta$ represents the vertical angle of the target's velocity. The radial velocity component is given by $V_R = \cos(\phi + \theta)V$.
   Then, the \gls{fim} can be reformulated as a function of $D$ and $V_R$:
   \begin{equation}\label{eq:dvfimafd}
    \mathbf{J}_{M}(D, V_R) =  -2\gamma
    \begin{bmatrix}
        \frac{\partial^2 \mathcal{A}\left(D,V_R\right)}{\partial D^2} & \frac{\partial^2 \mathcal{A}\left(D,V_R\right)}{\partial D \partial V_R} \\
        \frac{\partial^2 \mathcal{A}\left(D,V_R\right)}{\partial D \partial V_R} & \frac{\partial^2 \mathcal{A}\left(D,V_R\right)}{\partial^2 V_R}
    \end{bmatrix}_{\substack{D=0\\V_R=0}}
	\end{equation}
   Applying the chain rule we can get:
    \begin{align}
    &-\frac{\partial \mathcal{A}^2\left(D,V_R\right)}{\partial D^2}\bigg|_{\substack{D=0\\V=0}} = \frac{4f_s^2}{T^2c^2}\sum\limits_{n=1}^N \vert s[n] - s[n-1]\vert^2,\label{eq:dvfimdis11}\\
    &-\frac{\partial \mathcal{A}^2\left(D,V_R\right)}{\partial V_R^2}\bigg|_{\substack{D=0\\V_R=0}} = \frac{16\pi^2f_c^2}{c^2N^2}\sum\limits_{n=1}^Nn^2 \vert s[n]\vert^2,  
     \label{eq:dvfimdis12}\\
    \begin{split}
        &-\frac{\partial \mathcal{A}^2 \left(D,V_R\right)}{\partial D\partial V_R}\bigg|_{\substack{D=0\\V_R=0}} \\
    =&- \Im \left\{\frac{8f_sf_c\pi}{c^2NT}\sum\limits_{n=1}^N n(s[n] - s[n-1])s^*[n] \right\},
    \end{split}\label{eq:dvfimdis22}
	\end{align}
   where $\Im$ denotes the imaginary part of the expression. The \gls{crlb} can be defined as follows:
   \begin{equation}\label{eq:crlbn}
   \mathrm{CRB}(D) = \left[ \mathbf{J}^{-1}_{M}(D, V_R)\right]_{[1,1]}
   \end{equation}
   \begin{equation}\label{eq:crlbv}
   \mathrm{CRB}(V_R) = \left[ \mathbf{J}^{-1}_{M}(D, V_R)\right]_{[2,2]}
   \end{equation}
   The sqaure root of $\mathrm{CRB}(D)$ and $\mathrm{CRB}(V_R)$ represents the minimum possible variance in measuring $D$ and $V_R$. On the other hand, \gls{doa} estimation depends on neither directly. Instead, its lower bound is a monotonically decreasing function of \gls{snr}, as shown in \cite{crlbangularMark2021}. \revise{For simplicity, we use a fixed lower bound to represent the angle estimation error, as the $\mathrm{CRB}(\theta)$ is typically small and not the primary focus of this paper.}{}As introduced earlier, the spatial information of a pedestrian can be modeled as a random process $\mathbf{m}(t)$, where the trajectory and related quantities evolve over time. Each individual measurement of this process can be interpreted as a discrete sample drawn from the underlying random process. To represent the spatial information measured at the $k_{\mathrm{th}}$ time step, we define the corresponding vectors: ${\mathbf{m}}(k) = [{D}^k, {V_R}^k, {\phi}^k]^{{T}}$, ${\mathbf{e}}(k) = [e_D^k, e_{V_R}^k, e_{\phi}^k]^{{T}}$. Meanwhile, ${\mathbf{m}}(k)$ can be expressed as:
   \begin{align}
   &e_D^k \sim \mathcal{N}(0,\sigma_D^2),&\sigma_D = \sqrt{\mathrm{CRB}(D)};\label{eq:diserror}\\
   &e_{V_R}^k \sim \mathcal{N}(0,\sigma_{V_R}^2),&\sigma_{V_R} = \sqrt{\mathrm{CRB}(V_R)};\label{eq:vrerror}\\
   &e_{\phi}^k \sim \mathcal{N}(0,\sigma_{\phi}^2),&\sigma_{\phi} = \sqrt{\mathrm{CRB}(\phi)}.
   \label{eq:angleerror}
   \end{align}
   \subsection{Error model of data association}\label{subsec:errmodel}
   Unauthorized sensing likely operates in the \gls{ism} band to avoid legal consequences, while bandwidth can be limited. In indoor and urban environments, sensing is hindered by clutter from walls, buildings, and objects, resulting in distorted readings when tracking slow-moving pedestrians. While studies have shown micro-Doppler signatures can differentiate between pedestrians, cyclists, and vehicles \cite{doppler2018vander}, noisy signatures remain challenging. For data association, the \gls{jpda} algorithm effectively prunes infeasible hypotheses and calculates the most probable ones \cite{jpda2013mittal}.
   
   Drawing from these insights, we can postulate from the unauthorized sensing perspective: at any given moment, there exists a probability that the target's spatial information is not to be retrieved. While the interference and noise of the sensing initiator cannot be modeled, we can represent this uncertainty through two key factors: the target's radial velocity $V_R$ and the velocity resolution.
   The probability density function of measuring $V_R$ of the moving target can be described as:
   \begin{equation}
   P_t(V_R,\sigma_{V_R}) = \frac{1}{\sqrt{2\pi\sigma_{V_R}}}\exp{\left(-\frac{(\dot{V_R}-V_R)^2}{2\sigma_{V_R}^2}\right)}.
   \end{equation}  
   The cumulative difference between measuring static clutter and a moving target can be expressed using the error function as $p_k = \text{erf}\left(\frac{V_R}{\sqrt{2}\sigma_{V_R}}\right)$, where $\text{erf}(x) = \frac{2}{\sqrt{\pi}} \int_0^x e^{-t^2}dt$.
   This expression can also be extended to scenarios with multiple targets. We define a sample function $\delta(k)$ to denote whether a valid sample of $\mathbf{m}(t)$ can be obtained: 
   \begin{equation}\label{eq:meauncertain}
   \delta_k(k) = \begin{cases}
   1 & \text{with probability} \quad p_k ,\\
   0 & \text{with probability} \quad 1 - p_k.
   \end{cases}
   \end{equation}
   The sampled spatial information for a single target is then 
   \begin{equation}
       \mathbf{m}_{\sigma}(k) = \sum\limits_{k=1}^K \left(\mathbf{m}(k)+ \mathbf{e}(k)\right)\delta(k).
   \end{equation}
  
   \subsection{Error model of tracking performance}\label{subsec:perlow}
\begin{figure} 
    \centering
    \begin{subfigure}[b]{0.5\textwidth}
        \centering
        \includegraphics[width=0.45\textwidth]{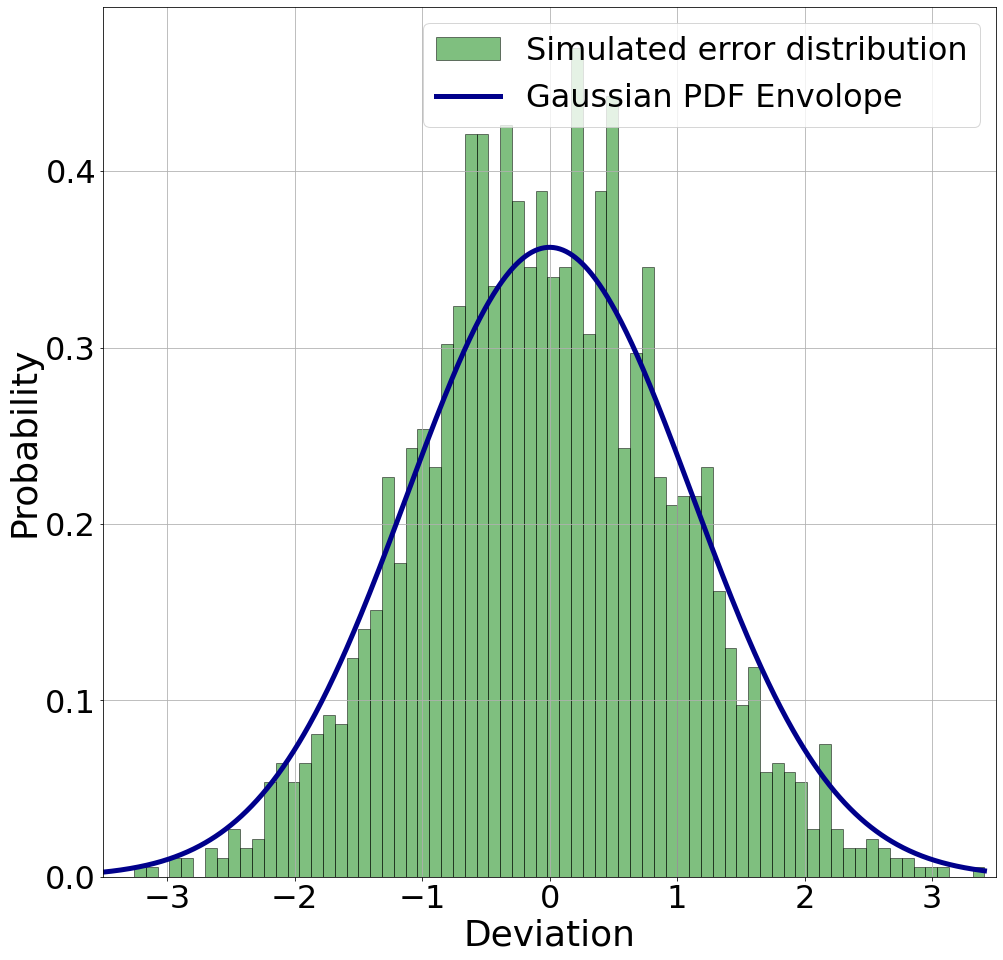}
        \caption{Simulated error compared to Gaussian PDF}
        \label{fig:trace_mani_bias}
    \end{subfigure}
    \hfill
    \begin{subfigure}[b]{0.24\textwidth}
        \centering
        \includegraphics[width=\textwidth]{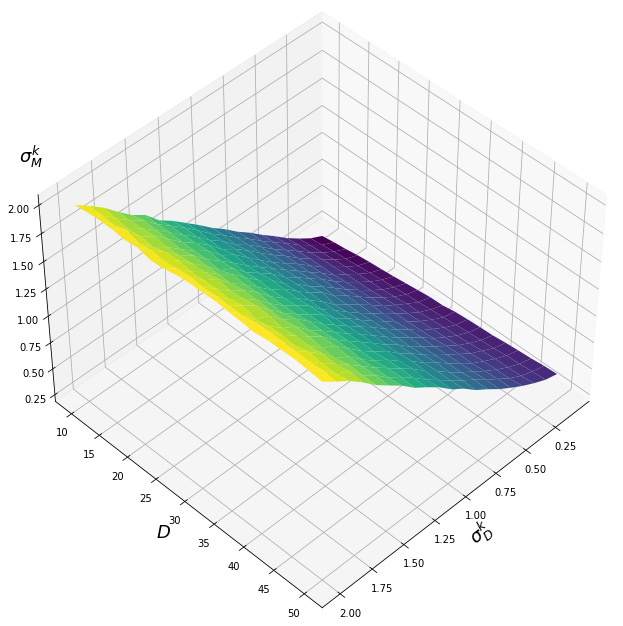}
        \caption{$\sigma_M^k$ w.r.t. $D$ and $\sigma_\phi^k$}
        \label{fig:trace_mani_ran}
    \end{subfigure}
    \begin{subfigure}[b]{0.24\textwidth}
        \centering
        \includegraphics[width=\textwidth]{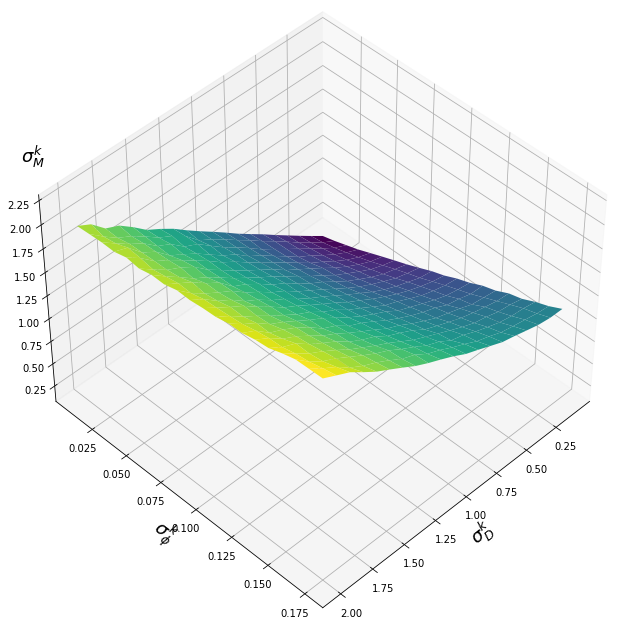}
        \caption{$\sigma_M^k$ w.r.t. $D$ and $\sigma_D^k$}
        \label{fig:trace_bia_coor}
    \end{subfigure}
    \vspace{-3mm}
    \caption{Error w.r.t. $D^k$, $\sigma_{\phi}^k$ and $\sigma_{D}^k$}
    \label{fig:taddemo}
   \end{figure}

   In a monostatic sensing scenario, the orientation of the target velocity usually can not be directly solved \cite{Orivelo2016Fair}. As depicted in Fig.~\ref{fig:monostatic}, the actual target velocity $V$ relates to radial velocity $V_R$ by $V = \frac{V_R}{\cos(\phi+\theta)}$, where $\phi$ is known and $\theta$ is unknown. We simplify tracking by omitting velocity estimation, as $V$ measurements are too noisy and uncertain for standard Kalman filtering.
   Using the cosine principle, the true position deviation between $\dot{\mathbf{m}}(k)$ and $\mathbf{m}(k)$ can be quantified as $\vert\Delta_D(k)\vert$, where
   \begin{align}
       \vert\Delta_D(k)\vert &= \sqrt{(D^k+e_D^k)^2 + (D^k)^2 -2D^k(D^k+e_D^k)\cos{e_{\phi}^k}}\nonumber\\
        &= \sqrt{2 (D^k + e_D^k) (1 - \cos{e_{\phi}^k}) + (e_D^k)^2}
   \end{align}
   The true position deviation $\vert\Delta_D(k)\vert$ can be assigned a sign based on the polarity of $e_D^k$, allowing it to be modeled as Gaussian distributed: $\Delta_D(k) \sim \mathcal{N}(0, \sigma_M^k)$. Simulation results in Fig.~\ref{fig:taddemo} indicate that the impact of $D^k$ on $\sigma_M^k$ is negligible when compared to the effects of $\sigma_D^k$ or $\sigma_{\phi}^k$. 
   
\begin{figure}[!htbp]
    \centering
   \includegraphics[width=0.37\textwidth]{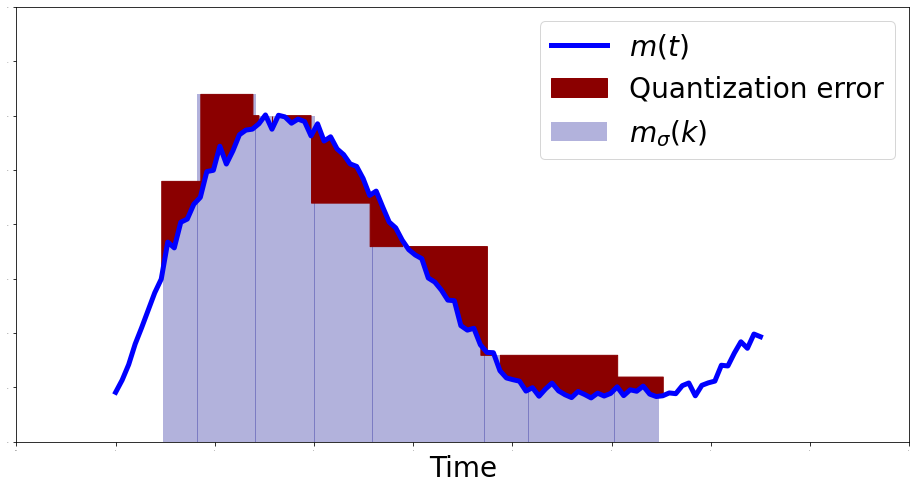}
   \vspace{-3mm}
   \caption{Error of non-uniform quantization}\label{fig:EqNQ}
   \end{figure}

   As introduced earlier, the target periodically measures the sensing capacity of the received signal. The sensing capacity is characterized by the probability that the sensing initiator retrieves a valid measurement within a time interval $ \Delta t $. For a longer interval $ \Delta T = K \Delta t $, the actual number of valid measurements $ \tilde{n} $ satisfies $ \tilde{k} \leqslant K $. Therefore, the tracking error can be evaluated in terms of the quantization error of this random process, depicted in Fig.~\ref{fig:EqNQ}. The average quantization interval can be determined by $\Delta q = \frac{K\Delta t}{\sum\limits_{k=1}^K p_k}$.
   With $\Delta q\vert_{K=1}= \frac{\Delta t}{p_1}$, the quantization error for data association is $\sigma_q^k = \frac{\Delta t^2}{12 p_k^2}$, derived from $\sigma_q = \frac{\Delta q^2}{12}$. Taking all possible errors into account, the performance lower bound can be expressed as a signal sampling error: $\sigma_p^k = \sigma^k_M + \sigma_q^k$, where $\sigma^k_M$ can be regarded as the signal noise power. 

   \section{Methodology}\label{sec:Meth}
   \subsection{Sensing components assessment}\label{subsec:scas}
   Sensing components typically satisfy two key requirements:\begin{enumerate*}[label=\emph{\roman*)}]
    	\item high correlation and 
    	\item low \gls{crlb} of sensing accuracy
    \end{enumerate*}. This stems from sensing initiators transmitting multiple pulses within an update interval to improve the \gls{snr} of target-reflected signals. These pulses are characteristically short, making them appropriate for short-range sensing applications. \revise{, and the signal is particularly susceptible to interference, especially in the \gls{ism} band where multiple devices coexist.}{} The high correlation between sensing pulses allows signal periodicity to be estimated by cross-correlating a short segment of the received signal with a slightly longer segment. Once periodicity is identified, shifted summation can be performed to enhance the \gls{snr}, allowing $\sigma_p^k$ to be accessed. Algorithm \ref{alg:csce} presents the detailed implementation of this approach. One concern arises from miss-detection of communication signals. While these signals can be highly correlated, they typically operate at lower power than sensing requires. Nevertheless, their distinctly higher \gls{crlb} makes them distinguishable and prevents triggering the mitigation strategies proposed in the following Subsection.
   \begin{algorithm}[!htbp]
    \caption{\gls{csce}}
    \label{alg:csce}
    \scriptsize
    \DontPrintSemicolon
    Input: received signal in an single assessment interval $r(n)$; short signal segment length $N_s$ and long signal segment length $N_l$\\
    \SetKwProg{Fn}{Function}{ :}{end}
    \Fn{\emph{CSCE}}{ 
        slice the received signal $r(n)$ with length $N_s$ and $N_l$ as $r_s$ and $r_k$\\
        $k = 0,1,2,\cdots , N_s + N_l -2 $\\
        $C(k) = \sum\limits_{l=1}^{N_l} r_s(l) r_l^*(l-k+N_l-1) $\\
        $C(k) = \frac{C(k)}{\max(C(k))} $\\
        $I \gets \{k : C(k) \text{ satisfies $C(k) > 0.707$}\}$\\ \tcp*{indices where the correlation is high} 
        $\Delta = \{I(i+1) - I(i) : i \in \{1,\ldots,|I|-1\}\}$\\ \tcp*{the difference of indices} 
        $N_{p} = |\{d \in \Delta : d \geqslant 0.4 \max(\Delta)\}|$\\ \tcp*{filter out closely located indices} 
        $ L = \left\lfloor\frac{\sum_{i=1}^{|\Delta|} \Delta_i}{N_{p}}\right\rfloor $ \\
        $ N_{all} = \frac{|r(n)|}{L_{period}} ; s(l) = \sum_{j=1}^{N_{all}} r(n - jL)$ \\ \tcp*{Shifted summation}
        $ h(l) = \frac{1}{100}; s_{ma}(l) = s(l)*h(l)$ \\ \tcp*{moving average}
        $ l_{start} = \min\{l : s_{ma}(l) \geqslant 0.707 \cdot \max(s_{ma})\}$\\
        $ l_{end} = \max\{l : s_{ma}(l) \geqslant 0.707 \cdot \max(s_{ma})\}$ \\ \tcp*{locate the sensing pulse}
        $p = s(l) $ for $ l_{start}\leqslant l \leqslant  l_{end}$\\
        estimate $\sigma_p^k$ based on $p$ and $\gamma$
        }
    \end{algorithm}
    \subsection{Privacy-persevering strategy}
     While $\sigma_p^k$ is estimated, the target can interfere with sensing by transmitting noise pulses in the direction of incoming sensing signals, thereby decreasing the \gls{snr} for the sensing initiator's receiver. This also exposes potential risks from the perspective of malicious actors, as distance estimation can occur while jamming power acquired. However, even if the sensing initiator estimates jamming power, distance calculations based on received jamming power remain highly unreliable due to severe path loss fluctuations. Additionally, transmission jamming power can be deliberately randomized to further mitigate this risk. Accordingly, we define two different monitoring strategies below to regulate the conditions under which jamming mode will be triggered.
   \subsubsection{Strategy I: Instant Monitoring}
   \begin{enumerate}
   \item[] \textbf{Step 1:} Monitor performance lower bound $\sigma_p(k)$. When $\sigma_p(t) < \theta_p$, increment $J_{count}$.
   \item[] \textbf{Step 2:} When $J_{count}$ reaches $\theta_J$, activate jamming mode: transmit noise pulse with length $L_j$ and power $P_j$, then reset $J_{count}$.
   \item[] \textbf{Step 3:} Return to \textbf{Step 1} after transmission.
   \end{enumerate}
  \subsubsection{Strategy II: Moving Average Monitoring}
  \begin{enumerate}
   \item[] \textbf{Step 1:} Monitor performance lower bound $\sigma_p(k)$. Calculate moving average $\bar{\sigma}_p = \frac{1}{K_m}\sum_{i=k-K_m+1}^{k}\sigma_p(k)$. When $\bar{\sigma}_p < \theta_p$, increment $J_{count}$
   \item[] \textbf{Steps 2 \& 3:} same as \emph{Strategy I}.
  \end{enumerate}
    
   \section{Simulation}\label{sec:simu}
   \subsection{Unauthorized sensing detection}

    Considering the $5.8$ GHz ISM band's higher center frequency and larger bandwidth compared to other ISM bands, factors that enhance sensing accuracy \cite{mahafza2005radar}, we assume unauthorized sensing occurs in this band. Furthermore, we employ \gls{lfm} pulses, favored in short-range applications through pulse compression \cite{mahafza2005radar}. The sensing pulse specifications are detailed in Table \ref{tab:spd}.
    
    After isolating the periodical component with Alg.~\ref{alg:csce}, we can use it to estimate $\mathrm{CRB}(D)$ and $\mathrm{CRB}(V_R)$. In practice, the target lacks knowledge of the reflected \gls{snr} in relation to the sensing initiator, and the actual sensing pulses are subject to noise and fading. Unlike conventional communication systems that use pilot signals for channel estimation to mitigate fading, blind channel estimation can be done without pilots but involves high computational complexity \cite{peken2017blind}. Therefore, we define the target-estimated $\mathrm{CRB}$ as $\mathrm{CRB}_\mathrm{T}$, and the actual $\mathrm{CRB}$ from the sensing initiator as $\mathrm{CRB}_\mathrm{I}$. To investigate the impact of fading channels, we evaluated the $\mathrm{CRB}_\mathrm{I}$ and $\mathrm{CRB}_\mathrm{T}$ through numerical simulations under different channel conditions. We considered three scenarios: \gls{awgn} channel, Rayleigh fading channel, and Rician fading channel with a K-factor of $2$. Besides, we consider equally powered noise and \gls{ofdm} symbols as noise and interference background. The $\mathrm{CRB}$ was then calculated using signal segments of $0.05\sec$.

    \begin{table}[!htbp]
		\centering
        \scriptsize
		\caption{Sensing pulse description}
		\label{tab:spd}
		\begin{tabular}{>{}m{0.2cm} | m{1.6cm} l m{3.7cm}}
			\toprule[2px]
			&\textbf{Parameter}&\textbf{Value}&\textbf{Remark}\\
			\midrule[1px]
			  &B &$100$ Mhz& Bandwidth \\

			&$f_c$ & $5.8$ Ghz& Carrier Frequency \\ \multirow{-2}{*}{\rotatebox{90}{\textbf{LFM}}}
			& $T_p$ & $0.1$ ms &  Pulse Duration\\
            & PRT & $0.4$ ms &  Pulse Repetition Time \\ 
            \midrule[1px]
		\end{tabular}
	\end{table}

      \begin{figure} 
    \centering
    \begin{subfigure}[b]{0.5\textwidth}
        \centering
        \includegraphics[width=0.49\textwidth]{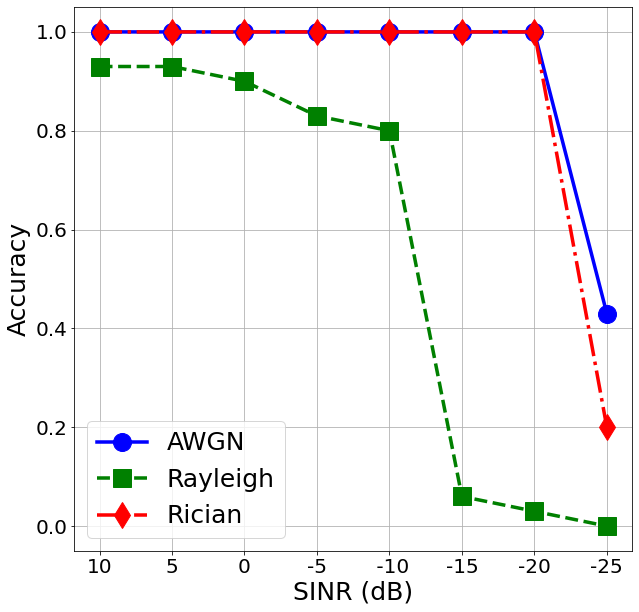}
        \caption{Sensing components extraction accuracy}
        \label{fig:acc_csce}
    \end{subfigure}
    \hfill
    \begin{subfigure}[b]{0.24\textwidth}
        \centering
        \includegraphics[width=\textwidth]{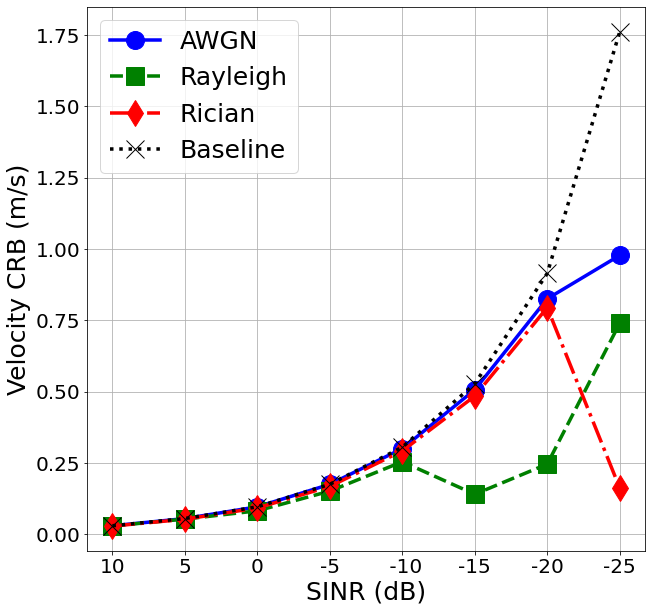}
        \caption{$\mathrm{CRB}(V_R)$}
        \label{fig:crlb_d}
    \end{subfigure}
    \begin{subfigure}[b]{0.229\textwidth}
        \centering
        \includegraphics[width=\textwidth]{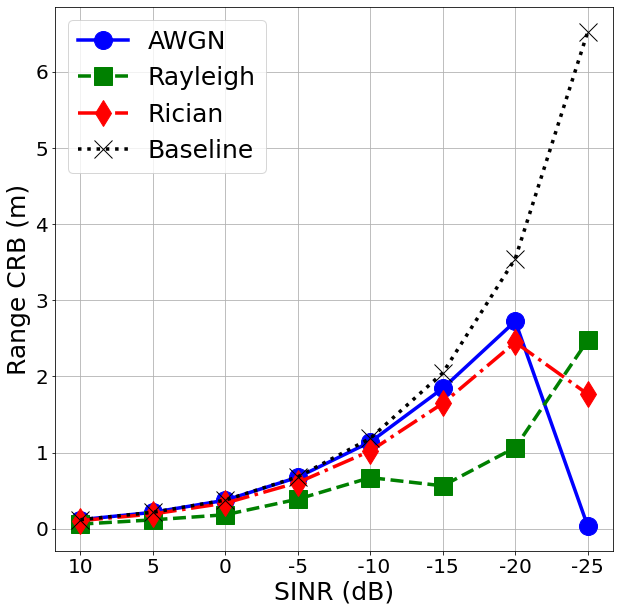}
        \caption{$\mathrm{CRB}(D)$}
        \label{fig:crlb_v}
    \end{subfigure}
    \caption{$\mathrm{CRB}(V_R)$, $\mathrm{CRB}(D)$ estimation and \gls{csce} accuracy under different channel response.}
    \label{fig:esti_acc}
   \end{figure}
    The simulation results are presented in Fig.~\ref{fig:esti_acc}. As shown in Fig.~\ref{fig:acc_csce}, under Rayleigh fading, when \gls{sinr} drops below $-10$ dB, \gls{csce} fails to detect sensing components, hence the estimated $\mathrm{CRB}$ cannot track the noise- and interference-free baseline ($\mathrm{CRB}_\mathrm{I}$). Meanwhile, such cutoff point exists when \gls{sinr} drops below $-20$ dB under \gls{awgn} and Rician fading. The Rayleigh fading channel introduces severe \gls{isi}, making sensing pulse detection more challenging. The \gls{isi} combined with noise results in estimated $\mathrm{CRB}$ values consistently below baseline, as evident in Figs.~\ref{fig:crlb_d}-~\ref{fig:crlb_v}, where $\emph{Baseline} > \emph{AWGN} > \emph{Rician} > \emph{Rayleigh}$ generally holds. These results indicate that the $\mathrm{CRB}$ may not be assessed when the $\mathrm{SINR}$ falls below $-10$ dB. Since reflected signals are significantly weaker than the target received signal, such weak signals become infeasible for sensing. Thus, the risk of being unable to access the $\mathrm{CRB}$ of sensing can be ignored. 
    \subsection{Unauthorized sensing mitigation}
    After evaluating the accuracy of the $\mathrm{CRB}$ in the previous subsection, we proceed to conduct further simulations that account for mobility and urban environments. For the urban path loss channel, we consider the \gls{3gpp} channel model under the \emph{Canyon Street} scenario \cite{3GPP}, examining both \gls{los} and \gls{nlos} conditions. For \gls{los} and \gls{nlos},
    \begin{align}\label{eq:los}
    \begin{split}
    L_{LOS} = 32&.4 + 21 \log_{10}(d) + 20 \log_{10}(f_c) + X_{\sigma};\\
    &X_{\sigma} \sim \mathcal{N}\left(0, 4 \right),
    \end{split} 
    \end{align}
    \begin{align}\label{eq:nlos}
    \begin{split}
      L_{NLOS} =& \;35.3 \log_{10}(d) + 22.4 + 21.3 \log_{10}(f_c) \\&- 0.3 (h-1.5) + X_{\sigma}; \\ &X_{\sigma} \sim \mathcal{N}\left(0, 7.82 \right).  
    \end{split}
    \end{align}
    Where $d$ denotes the distance between the transmitter and receiver. For pedestrians on the ground, we assume a height of \SI{1.5}{\meter}. To simplify calculation of reflected signal path loss, we approximate it by using twice the direct distance, neglecting signal absorption effects. The channel response is then modeled using Rayleigh fading for \gls{nlos} conditions and Rician fading for \gls{los} conditions.

\begin{figure}[!htbp]
    \centering
   \includegraphics[width=0.41\textwidth]{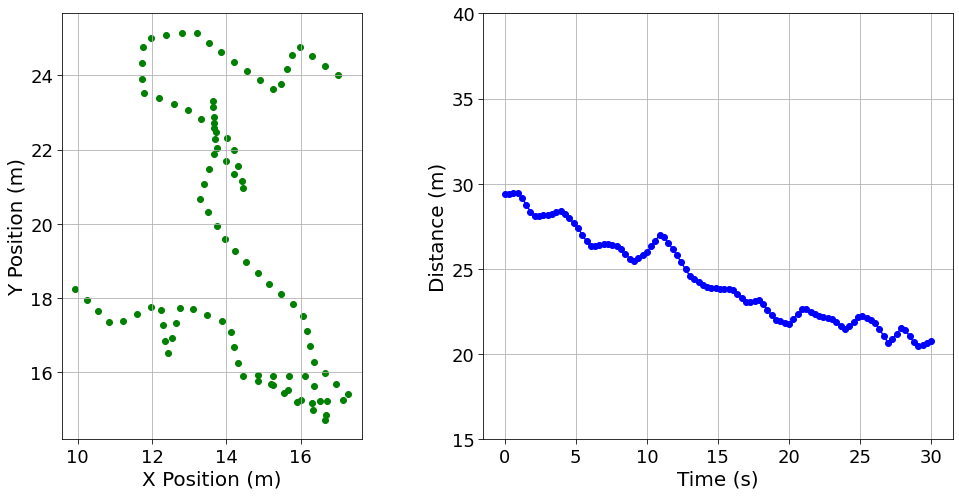}
   \caption{Randomly generated trajectory}\label{fig:ran_traj}
   
   \end{figure}

    For mobility modeling of a pedestrian, we consider an average walking speed of \SI{1.4}{\meter/\second} with directional movements subject to confined randomness. The constrained random trajectory of a pedestrian's movement pattern is illustrated in Fig.~\ref{fig:ran_traj}. As the pedestrian moves through the urban environment, the channel conditions alternate between \gls{los} and \gls{nlos}, simulating realistic radio propagation characteristics. Since reflected signals are weaker than those received directly by the target, we compensate by adding $\beta_r$ to the estimated $\mathrm{SINR}$, where $\beta_r$ is derived from the average difference between Eqs.~(\ref{eq:los}) and (\ref{eq:nlos}) at distances of \SI{25}{\meter} and \SI{50}{\meter}.
 
      \begin{table}[!htbp]
		\centering
        \scriptsize
		\caption{Simulation Setup I}
		\label{tab:sys1}
		\begin{tabular}{>{}m{0.15cm} | m{1.2cm} l m{3.6cm}}
			\toprule[2px]
			&\textbf{Parameter}&\textbf{Value}&\textbf{Remark}\\
			\midrule[1px]
			  &B &$50,150$ Mhz& Bandwidth \\
			&$T_r$ & $15$ dBm & Transmitting power \\ \multirow{-2}{*}{\rotatebox{90}{\textbf{System}}}
			& $N_f$ & $-92$ dBm & Noise floor in ISM band \cite{noisefH2017}\\
            & $\mathrm{CRB}(\phi)$ & $0.02$ rad & Angle estimation lower bound \\ 
            & $\beta_r$ & $-6$ dB & Compensation factor for reflection \\
            \midrule[1px]
		\end{tabular}
	\end{table}
    \begin{figure} 
    \centering
    \begin{subfigure}[b]{0.40\textwidth}
        \centering
        \includegraphics[width=\textwidth]{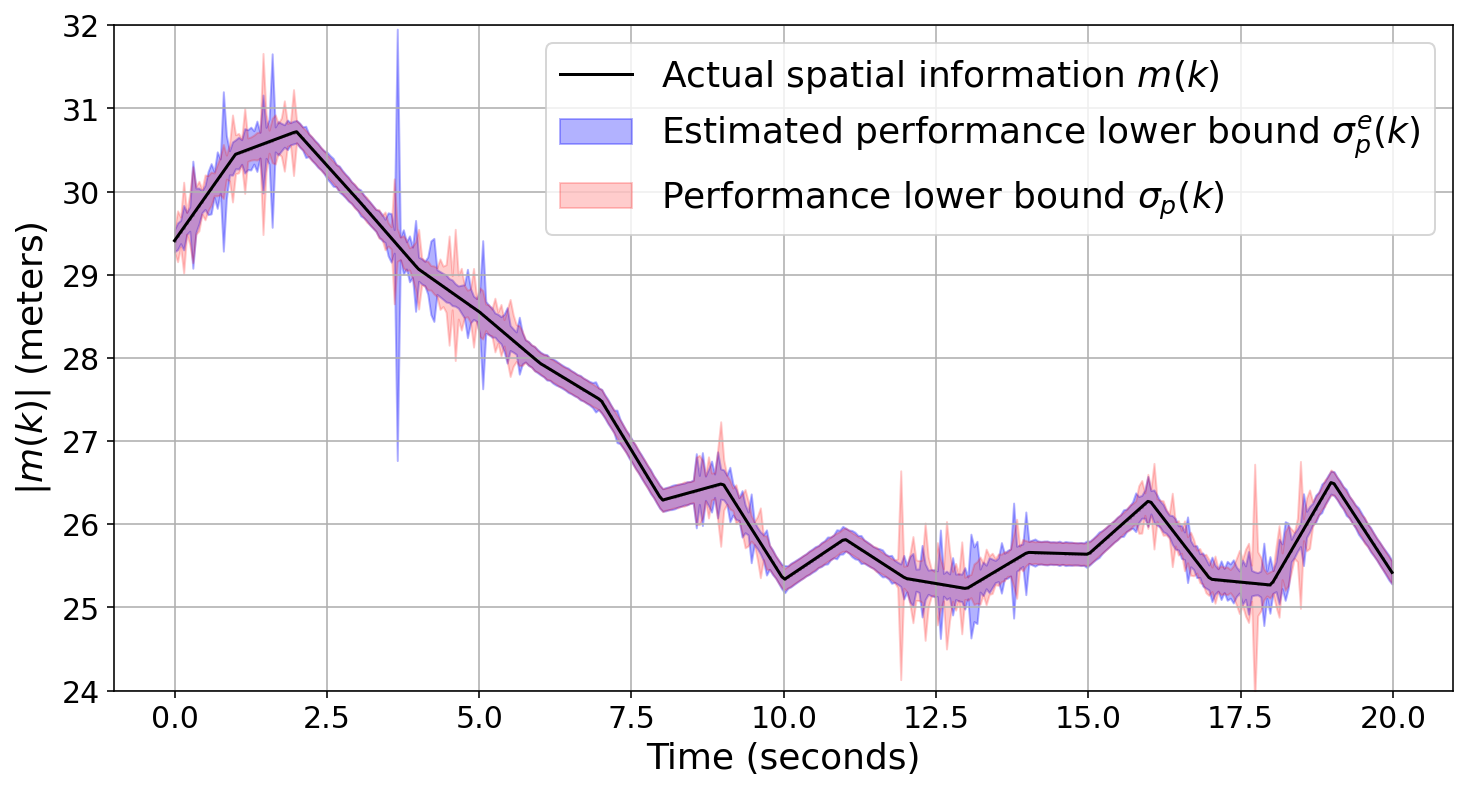}
        \caption{$150$ Mhz bandwidth}
        \label{fig:50MHZ}
    \end{subfigure}
    \hfill
    \begin{subfigure}[b]{0.40\textwidth}
        \centering
        \includegraphics[width=\textwidth]{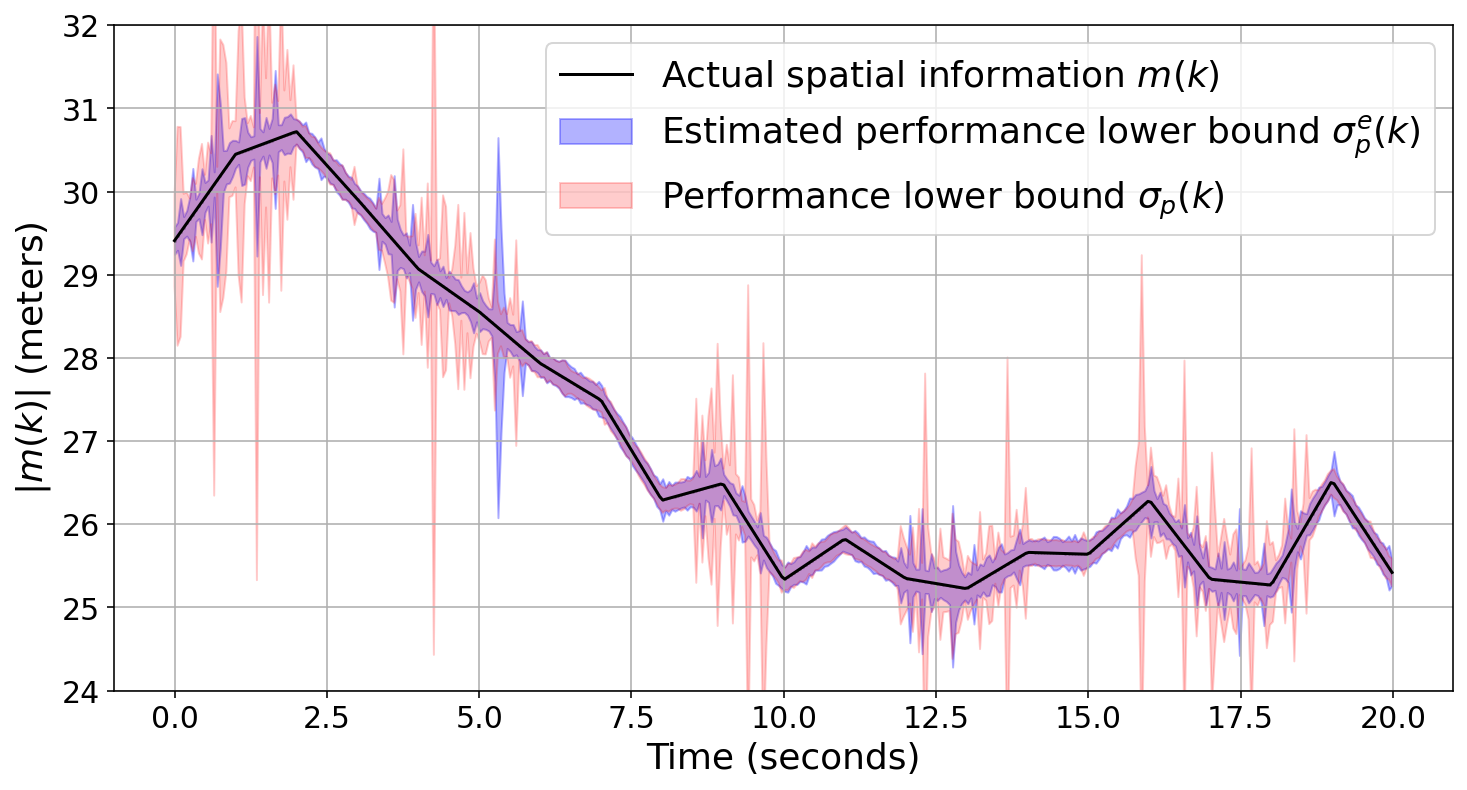}
        \caption{$50$ Mhz bandwidth}
        \label{fig:150MHZ}
    \end{subfigure}
    \caption{Performance lower bound of \gls{lfm} pulses with three different bandwidths}
    \label{fig:PLBM}
   \end{figure}

   Simulations were conducted across two bandwidths using the parameters in Tab.~\ref{tab:sys1}, assuming negligible external interference. Fig.~\ref{fig:PLBM} shows the results,  with the performance lower bound visualized as a symmetric shaded region around $\mathbf{m}(k)$. The results show that performance lower bounds are more accurately described when they are small. The average actual lower bounds for two bandwidths are \SI{0.20}{\meter} and \SI{0.43}{\meter}, compared to estimated values of \SI{0.17}{\meter} and \SI{0.24}{\meter}. This discrepancy arises because signals with larger bandwidths are more robust to multi-path effects.
    
\begin{figure}
    \begin{subfigure}[b]{0.40\textwidth}
        \centering
        \includegraphics[width=\textwidth]{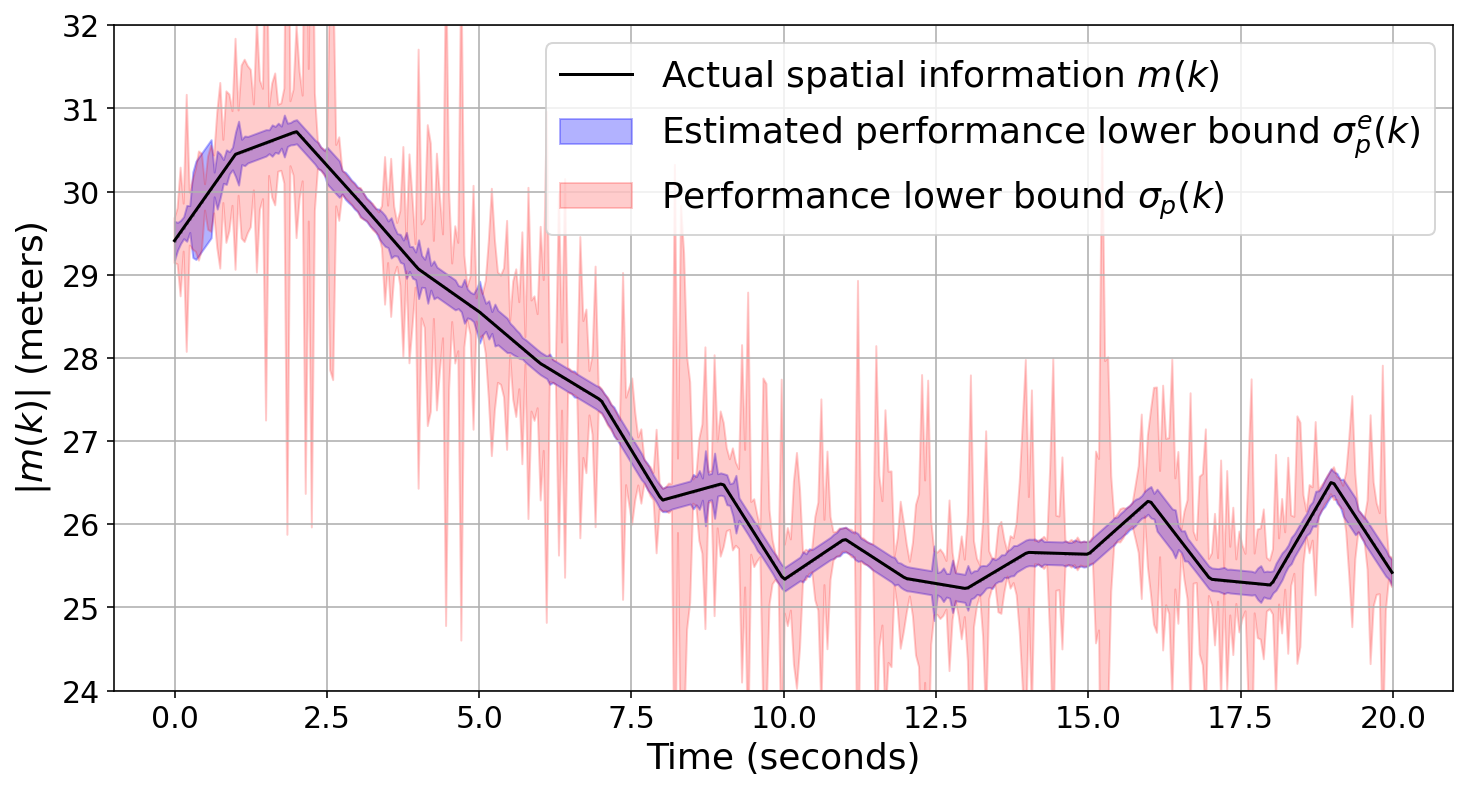}
        \caption{$50$ Mhz bandwidth}
        \label{fig:j50MHZS1}
    \end{subfigure}
    \hfill
    \begin{subfigure}[b]{0.40\textwidth}
        \centering
        \includegraphics[width=\textwidth]{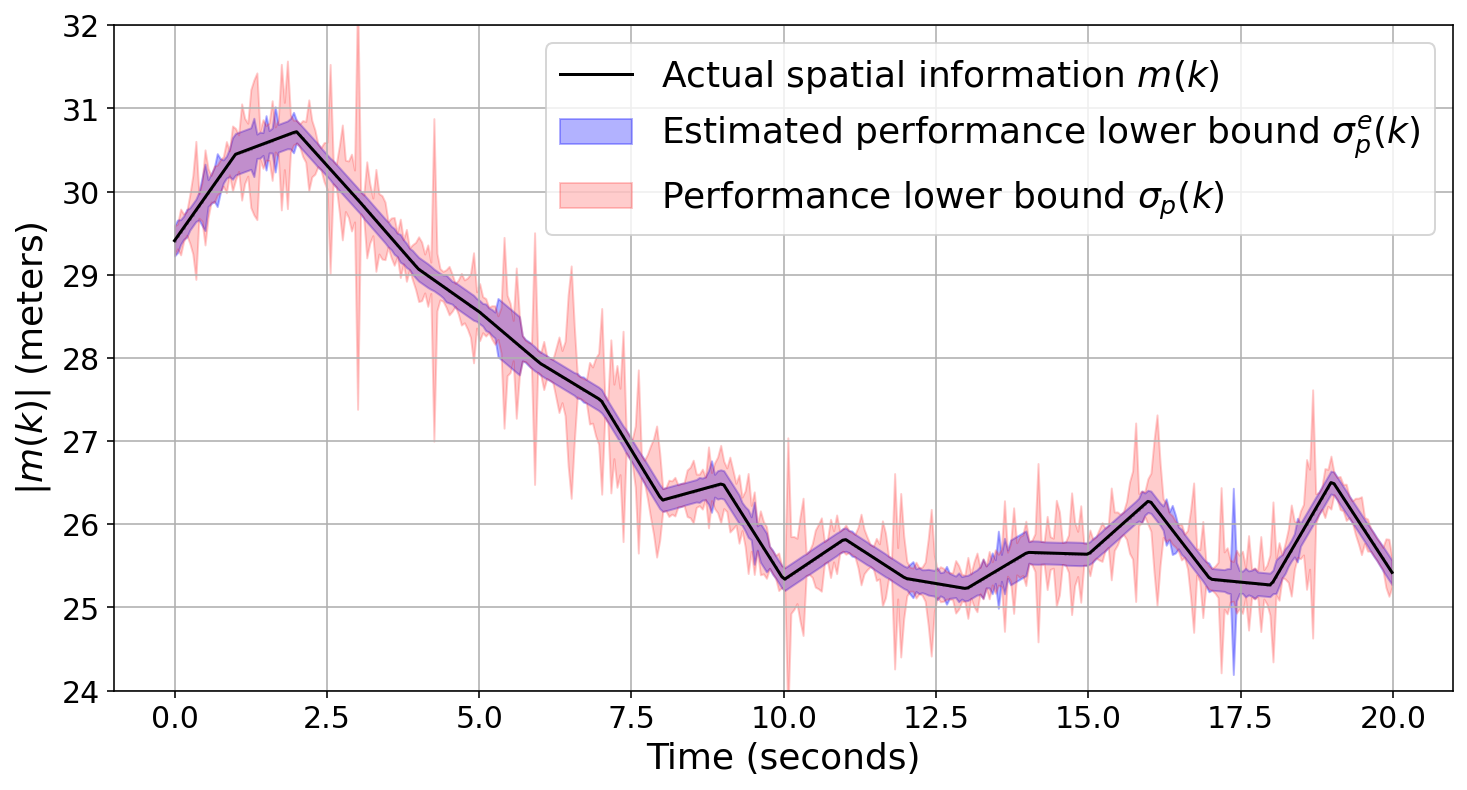}
        \caption{$150$ Mhz bandwidth}
        \label{fig:j150MHZS1}
    \end{subfigure}
    \caption{Performance lower bound under \emph{Strategy I}}
     \label{fig:JPLBM1}
    \end{figure}

\begin{figure}
    \begin{subfigure}[b]{0.40\textwidth}
        \centering
        \includegraphics[width=\textwidth]{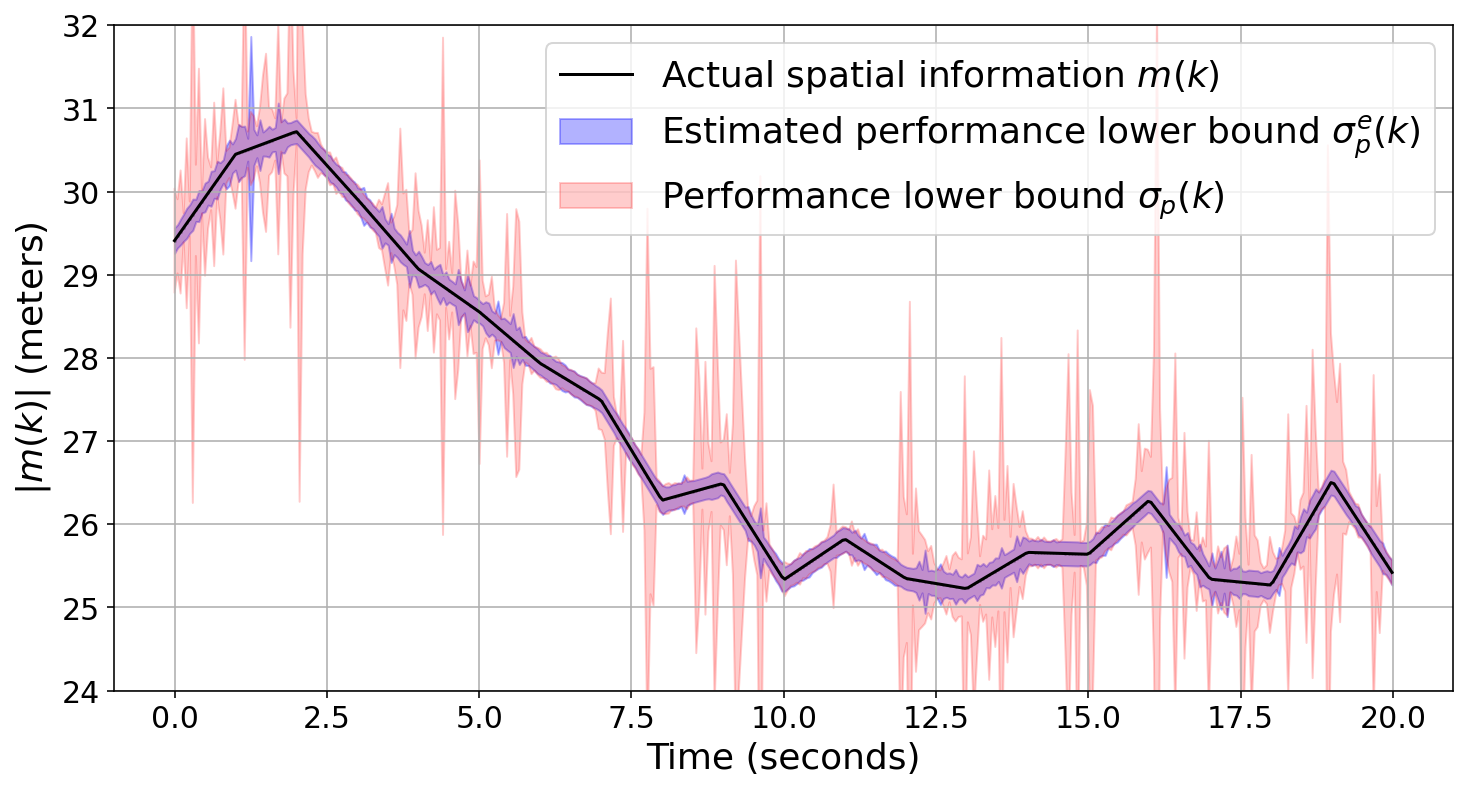}
        \caption{$50$ Mhz bandwidth}
        \label{fig:j50MHZS2}
    \end{subfigure}
    \begin{subfigure}[b]{0.40\textwidth}
        \centering
        \includegraphics[width=\textwidth]{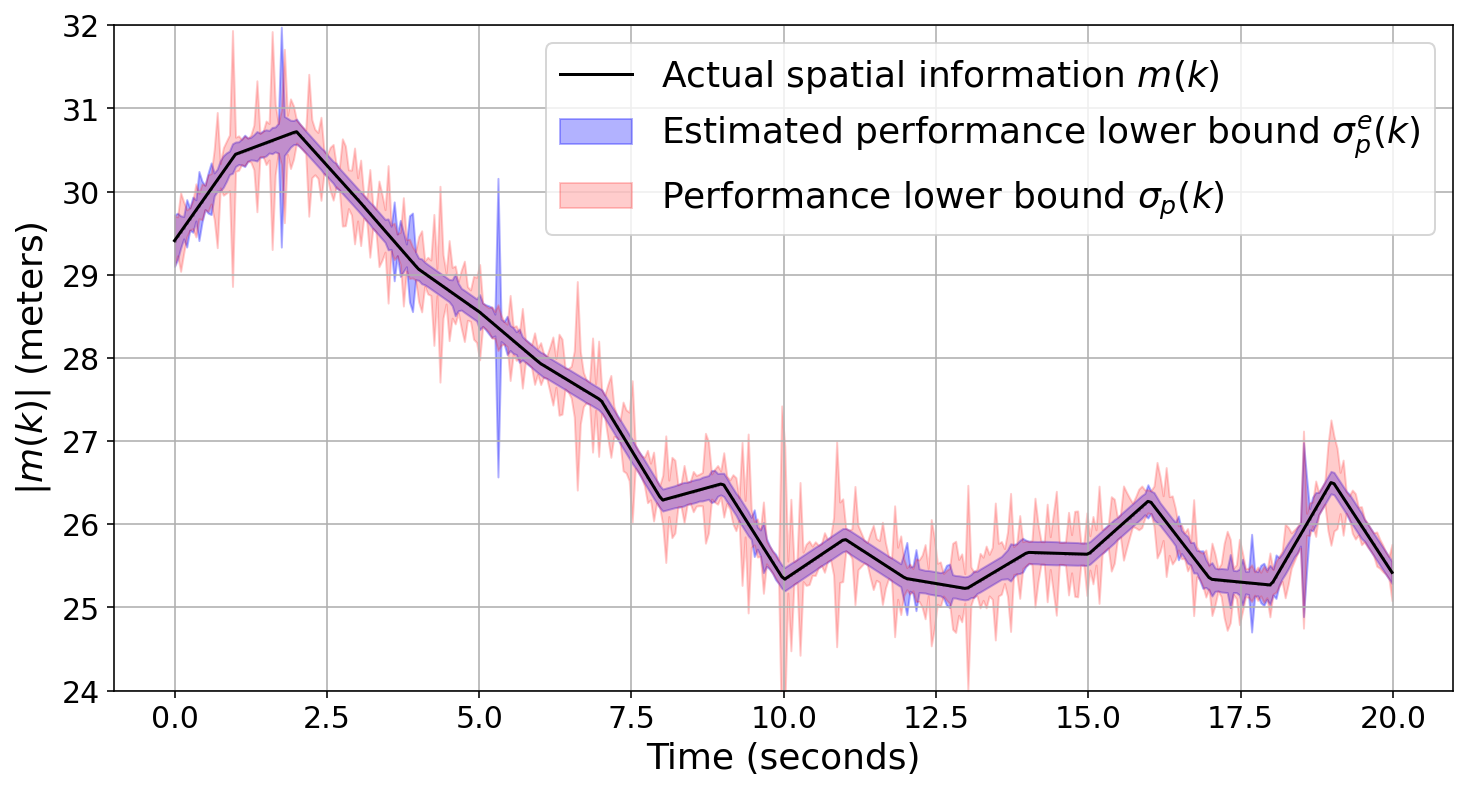}
        \caption{$150$ Mhz bandwidth}
        \label{fig:j150MHZS2}
    \end{subfigure}
    \caption{Performance lower bound under \emph{Strategy II}}
    \label{fig:JPLBM2}
   \end{figure}

   We validate the proposed monitoring strategies against unauthorized sensing systems through simulations using two sensing pulses with bandwidths of \SI{50}{\mega\hertz} and \SI{150}{\mega\hertz}. The jamming noise power is set as $P_j = a_j P_r$, where $P_r$ denotes the received sensing signal power at the target, and other simulation parameters are listed in Tab.~\ref{tab:sys2}. The simulation results, presented in Figs.~\ref{fig:JPLBM1} and \ref{fig:JPLBM2}, demonstrate a significant increase in $\sigma_p(k)$ when compared to the results shown in Fig.~\ref{fig:PLBM}. This increase is particularly pronounced in the "smooth" regions of Fig.~\ref{fig:PLBM}, where \gls{los} exists between the target and sensing initiator. Quantitatively, under \emph{Strategy I}, the mean of $\sigma_p(k)$ increased from \SI{0.43}{\meter} to \SI{0.78}{\meter} for  \SI{50}{\mega\hertz} bandwidth, and from \SI{0.20}{\meter} to \SI{0.33}{\meter} for \SI{150}{\mega\hertz} bandwidth. Over 10 simulations, the jamming mode was triggered on average 35 times for \SI{50}{\mega\hertz} bandwidth and 50 times for \SI{150}{\mega\hertz} bandwidth. Under \emph{Strategy II}, the mean of $\sigma_p(k)$ increased to \SI{0.52}{\meter} for \SI{50}{\mega\hertz} bandwidth signals and \SI{0.32}{\meter} for \SI{150}{\mega\hertz} bandwidth, with average trigger counts of 20 and 45 times, respectively. These simulation results show \emph{Strategy I} achieves higher sensitivity and stronger interference with unauthorized sensing systems, though \emph{Strategy II} triggers less frequently—particularly for \SI{50}{\mega\hertz} bandwidth signals. Since \SI{50}{\mega\hertz} bandwidth signals already exhibit large $\sigma_p(k)$, \emph{Strategy II}'s reduced trigger frequency offers a more energy-efficient solution for mobile devices while maintaining effective monitoring performance.
  \begin{table}[!htbp]
		\centering
        \scriptsize
		\caption{Simulation Setup II}
		\label{tab:sys2}
		\begin{tabular}{>{}m{0.15cm} | m{1.2cm} l m{3.6cm}}
			\toprule[2px]
			&\textbf{Parameter}&\textbf{Value}&\textbf{Remark}\\
			\midrule[1px]
			  &$\theta_p$ &\SI{0.17}{\meter}& Performance threshold \\
			&$\theta_J$ & $3$ & Count number threshold \\ \multirow{-2.5}{*}{\rotatebox{90}{\textbf{Mitigation}}}
			& $L_j$ & $\sim\mathcal{U}(50,350)$ ms & Noise pulse duration\\
            & $a_j$ & $\sim\mathcal{U}(-3,3)$ dB & Power modifier \\ 
            & $K_m$ & $3$ & Steps of moving average \\
            \midrule[1px]
		\end{tabular}
	\end{table}
   \section{Conclusion}\label{sec:conclu}
  This paper presents a framework enabling individuals to monitor unauthorized sensing activities in \gls{ism} bands. We investigated accessing $\mathrm{CRB}$ in urban environments without computationally expensive blind channel estimation. The accessed $\mathrm{CRB}$ was modeled as a random process with sampling and quantization errors to characterize tracking accuracy of individuals. Next, we proposed and validated a jamming strategy that reduces tracking accuracy to preserve privacy. Given the proliferation of sensing techniques, this work offers crucial insights for preserving individual privacy in increasingly monitored environments.

    
\bibliographystyle{IEEEtran}
\bibliography{references}

\end{document}

%% file: glossary.tex
\makeglossaries
\newacronym{rgd}{RGD}{robust gradient descend}
\newacronym{mle}{MLE}{maximium likehood estimation}
\newacronym{spgd}{SPGD}{sample pruning gradient descend}
\newacronym{af}{AF}{ambiguity function}
\newacronym{jcas}{JCAS}{Joint Communication and Sensing}
\newacronym{fmcw}{FMCW}{Frequency Modulated Continuous Wave}
\newacronym{stft}{STFT}{Short Time Fourier Transform}
\newacronym{snr}{SNR}{Signal to Noise Ratio}
\newacronym{fim}{FIM}{Fisher Information Matrix}
\newacronym{crlb}{CRB}{Cram\'er-Rao bound}
\newacronym{doa}{DOA}{Direction of Arrival}
\newacronym{md}{MD}{micro-doppler}
\newacronym{jpda}{JPDA}{Joint Probability Data Association}
\newacronym{ism}{ISM}{Industrial, Scientific, and Medical}
\newacronym{los}{LOS}{Line of Sight}
\newacronym{nlos}{NLOS}{Non-line of Sight}
\newacronym{csce}{CSCE}{Correlation-based Sensing Components Extraction}
\newacronym{awgn}{AWGN}{additive white Gaussian noise}
\newacronym{ofdm}{OFDM}{orthogonal frequency-division multiplexing}
\newacronym{sinr}{SINR}{signal to interference and noise ratio}
\newacronym{isi}{ISI}{inter-symbol interference}
\newacronym{3gpp}{3GPP}{\emph{3rd Generation Partnership Project}}
\newacronym{lfm}{LFM}{Linear Frequency Modulation}
\newacronym{csi}{CSI}{Channel State Information}
\newacronym{aoa}{AOA}{Angle of Arrival}
\newacronym{snir}{SNIR}{Signal to Noise and Interference Ratio}